\documentclass[final,5p,times,twocolumn]{elsarticle}

\usepackage{amsmath,amssymb}

\usepackage{aas_macros}
\usepackage[utf8]{inputenc}
\usepackage{graphicx}
\usepackage{dcolumn}
\usepackage{bm}
\usepackage{xcolor}
\usepackage{stfloats}
\usepackage{caption}
\usepackage{gensymb}
\usepackage{natbib}

\usepackage{subcaption} 
\usepackage{multirow}
\usepackage[colorlinks=true, allcolors=blue]{hyperref}
\usepackage{soul}
\begin{document}

\begin{frontmatter}

\title{Constraints on GRB Jet Properties from IceCube Upper Limits: Insights from GRB~221009A and GRB~240825A}

\author[inst1]{Chiranjeet Pradhan}
\author[inst2,inst3]{Khushboo Sharma}
\author[inst41,inst42]{Abhijit Roy}
\author[inst2,inst5]{Jagdish C. Joshi \corref{cor1}}
\ead{jagdish@aries.res.in}
\cortext[cor1]{Corresponding author}
\address[inst1]{Sardar Vallabhbhai National Institute of Technology (SVNIT), Ichchhanath, Surat 395007, India}
\address[inst2]{Aryabhatta Research Institute of Observational Sciences (ARIES), Manora Peak, Nainital 263001, India}
\address[inst3]{Department of Physics, Indian Institute of Technology Roorkee, Roorkee 247667, India}
\address[inst5]{Centre for Astro-Particle Physics (CAPP) and Department of Physics, University of Johannesburg, PO Box 524, Auckland Park 2006, South Africa}

\address[inst41]{Gran Sasso Science Institute, via F. Crispi 7 -- 67100, L'Aquila, Italy} 
\address[inst42]{INFN/Laboratori Nazionali del Gran Sasso, via G. Acitelli 22, Assergi (AQ), Italy} 

\begin{abstract}
The IceCube neutrino telescope has provided upper limits on neutrino emission from gamma-ray bursts. These constraints provided by the IceCube detector have been instrumental in investigating the properties of the GRB jet and its emission models. During the prompt phase of gamma-ray burst emission, intense radiation components are generated that interact with the shock-accelerated particles within the jet. We study various GRB emission models, such as the internal shock model, the photospheric models, and also include a model-independent case. Based on these models, we calculate the neutrino fluence using the photo-hadronic interaction process. We estimate the bulk Lorentz factor using the well-known correlations between prompt phase observables, which is then used to calculate the emission site for the model-dependent scenarios. For GRB~221009A, we find that a low baryon loading scenario is consistent with the IceCube upper limits; however, for GRB~240825A, a higher value of baryon loading is preferred. Also, the values of the microphysical parameters $\epsilon_e$ and $\epsilon_B$ for GRB~240825A are lower by factors of approximately 10 and 100, respectively, compared to those of GRB~221009A. Further, using neutrino upper limits for these two sources, we estimate the lower limits on the dissipation radius for our models. The current TeV–PeV upper limits for GRB~221009A are already useful for constraining parameter space for the BPH and MPH models.
\end{abstract}


\end{frontmatter}

\section{Introduction}

The IceCube detector \citep{karle2010icecube}, located at the South Pole, has been very useful in investigating the non-thermal particle population in gamma-ray bursts (GRBs) and their radiation mechanisms \citep{2017ApJ_Artsen}. Since May 2011, neutrino events have been recorded in this detector with the full capacity of the 86-string configuration distributed in a volume of $\sim 1~\mathrm{km^{3}}$ \citep{Aartsen2013ApJ.._132A}. Another cubic kilometre neutrino telescope, located in the Northern Hemisphere, KM3NeT, has been operating in partial mode since 2016 \citep{Adri2016JPhG_01A,sinu2021JInst_S}. Before KM3NeT, its pilot project ANTARES telescope, with 12 vertical strings distributed in $\sim 0.01~\mathrm{km^{3}}$, reported the neutrino upper limits in the multi-TeV range from some of the very high energy GRBs such as GRB~180720, GRB~190114C, and  GRB~190829A \citep{antares2021JCAP2A, Aiello2024JC_06A}. However, due to its lower effective area, these upper limits are less restrictive for studying the properties of the GRB jet. Recently, KM3NeT reported the neutrino upper limits for the very high energy GRB~221009A using a 21-string configuration in its ARCA detector and 10-strings in its ORCA detector \citep{Aiello2024JC_06A}. However, these limits in the multi-TeV range are still higher by an order of magnitude compared to the IceCube upper limits; hence, we will use the IceCube upper limits in our study to constrain the prompt emission models. So far, the IceCube detector's spatial and temporal searches of neutrino events from GRBs have yielded a null detection \citep{abbasi2012Natur_I, Aartsen2015ApJ_5A,aartsen2016ApJ_115A, Abbasi2022ApJ_116A}. This is in contrast to the predicted rate of $\sim 10 - 100$ neutrino events per year from GRBs by a $\rm km^3$ detector \citep{waxman1997PhRvL2W}. The neutrino signal from the GRBs is instrumental in understanding the hadronic content (such as protons and heavy nuclei) of the jet \citep{wang2008ApJ_2W,jagdish2016_9J}. The particle acceleration mechanisms, such as diffusive shock acceleration \citep{mgbaring2011AdSpRB}, stochastic acceleration \citep{asano2015MNRAS_2242A}, or magnetic reconnection \citep{zhang2011ApJ_90Z, Peng_2024ApJ_62P}, can generate a population of non-thermal particles at the shocks inside the GRB jet. Interactions of these relativistic particles with the radiation field inside the jet lead to the production of secondary gamma-rays and neutrino fluxes \citep{waxman1997PhRvL2W, Murase_2006, Asano2014ApJ_54A, Sahu2024Ap_224S, sahu2024MNRA_64S}.

In recent years, IceCube follow-up observations of individual GRBs have been extensively used to investigate their jet properties and emission models \citep{He2012ApJ_.29H, Mei2022ApJ_82M, Liu2023ApJ_2L, Veres_2024}. For example, GRB~080319B located at a redshift $z \sim 0.937$, was very bright in optical, and during its prompt phase, the released isotropic equivalent gamma-ray energy was $E_{\gamma, \rm iso} \simeq 1.3 \times 10^{54}$ erg, in the range of 20 keV to 7 MeV \citep{racusin_2008Natur_R}. IceCube, using its initial configuration of 59 strings \citep{abbasi2009ApJ_1A}, followed this GRB, and considering the internal shock (IS) model for GRB prompt radiation,  the prompt phase neutrino spectrum was also calculated. This model predicted that the neutrino event from this source would be $\sim 0.1$ for a bulk Lorentz factor of $300$. Another very luminous GRB~130427A emitted $E_{\gamma, \rm iso} \simeq 1.4 \times 10^{54}$ erg \citep{Ackermann2014Sci_42A} of gamma-ray energy, located at a redshift of value $z  \sim 0.34$ \citep{levan2013GCN_1L}, was followed by the IceCube using their full 86-string configuration, and null detection was reported \citep{Blaufuss2013GCN_1B}.  These non-detection results were used by \citep{Gao_2013} to study the jet parameters, such as emission radius, bulk Lorentz factor, fraction of energy available in protons, etc. They reported that in a model-independent case, the emission radius is inversely proportional to the bulk Lorentz factor. In the case of model-dependent scenarios, they discussed the IS, the baryonic photospheric model (BPH), and the magnetic photospheric model (MPH). In the case of the MPH model, they found a restrictive constraint that the ratio between the fraction of energy in protons and electrons was less than 2 and approximately independent of the bulk Lorentz factor. 

The low isotropic gamma-ray energy $E_{\gamma, \rm iso} \simeq 4 \times 10^{52}$ erg of another GRB~230307A, which is located at a redshift $z = 0.065$ \citep{levan2024Natur_737L}, also provided upper limits on the neutrino signal \citep{IC2023GCN_1I}. Using the photo-hadronic scenario of neutrino production, \citep{Song2023ApJ_133S} found that the prompt radiation mechanism and baryon loading cannot be constrained. However, this long GRB was produced via a binary compact star merger \citep{levan2024Natur_737L, Wng2024ApJ_.9W}.
Hence, hadronuclear (neutron-proton, proton-proton, and neutron-neutron) interactions are expected in a neutron-rich jet medium. These interactions can produce a quasi-thermal neutrino flux in the GeV to sub-TeV range \citep{Koers2007AnA._95K}. \citep{Song2023ApJ_133S} studied a hadronuclear interaction model and constrained the jet composition of GRB~230307A using IceCube upper limits. They reported that the baryon loading in GRB~230307A is roughly an order of magnitude larger than that in GRB~221009A.

In this work, we examine two long GRBs, GRB~221009A and GRB~240825A, which have comparable neutrino fluence upper limits \citep{IC2024GCN.37326....1I}. Both are produced by massive-star core collapse \citep{Blanchard2024NatA_4B, Gupta2025_00142G}; however, they have distinct prompt emission characteristics as listed in Table \ref{tab:GRBP}. We consider various emission models for the prompt phase of GRBs and explore their parameter space using neutrino fluence upper limits.

The outline of this paper is the following. In the Subsection below, we describe the internal shock and photospheric radius, and the definition of the Lorentz factor used. We also describe the neutrino fluence based on the photo-hadronic interaction model. We discuss the model parameter space for GRB~221009A and GRB~240825A by using IceCube upper limits in Section \ref {sec:case_study}. In Section \ref {sec:Optical_depth}, we check the pair production opacity for maximum energy photons in the prompt phase, and estimate the $R-\Gamma$ parameter space for the gamma-ray and neutrino emission. We summarize and discuss our results in Section \ref{sec:results}.

\subsection{Prompt Emission Site and Photo-hadronic Interactions in the GRB Jet}
\label{IS-PH_Model}

During the prompt phase of the GRBs, gamma-ray spectra can be explained via the IS model \citep{Rees1994ApJ_93R,daigne1998MNRAS_5D} or via photospheric dissipation models \citep{Pacz1990ApJ_218P, Thomp1994MNR_80T, mesza2000ApJ_292M, gia2005AG, Rees2005ApJ_847R, Pe_er_2007, Beloborodov_2010}. The fireball kinetic energy dissipates at the IS regions due to collisions between relativistic shells produced by the central engine. In these shocks, electrons and nuclei gain energy and also radiate non-thermal radiation. The IS dissipation radius can be related to the minimum variability time scale (\( \delta t_{\rm var} \)), which reflects the shortest ejection timescale of the central engine. In this case, the emission radius is $R_{\rm IS} \approx 2 \Gamma^{2} c \delta t_{\rm var}$ \citep{Rees1994ApJ_93R} and $\Gamma$ is the bulk Lorentz factor of the outflow at the IS.

In the photospheric model, gamma-ray photons are released close to or just above the photosphere when the Thompson optical depth for photon-electron interactions is approximately equal to unity, i.e., $\tau_{\gamma e} \simeq 1$ \citep{Giannios_2008, Veres_2012, Beniamini_2017}. However, these radiations are produced closer to the central engine, where a higher density of $e^{\pm}$ pairs, electromagnetic radiation and baryons leads to the thermalization of this radiation \citep{Rees2005ApJ_847R}. In the case of a magnetically dominated jet, it is termed the case of Poynting flux-dominated or the MPH model, in which the dissipation mechanism, such as magnetic reconnection or kink instability, accelerates the particles, such as electrons, and their cooling via the synchrotron mechanism produces the radiation \citep{Thomp1994MNR_80T,gia2006A_..887G}.  In the BPH scenario, the jet is matter-dominated, and the energy is carried mostly by baryons. The dissipation of this energy leads to the acceleration of electrons. Further, the Comptonization process takes place and generates a broadened quasi-thermal spectrum \cite{rees2005ApJ..847R,Rees_2005}. The dissipative radius in these two cases is defined as \citep{Veres_2012, Meszaros_2000}

\begin{eqnarray} \label{R_ph}
    R_{\text{PH}} = r_{0} \, \eta_{T}^{1/\mu}
    \begin{cases}
    \left( \dfrac{\eta_{T}}{\eta} \right)^3~~~~~~~~~~, & \eta < \eta_{T} \  \\
    \left( \dfrac{\eta_{T}}{\eta} \right)^{1/(1 + 2\mu)}, & \eta > \eta_{T} \ 
    \end{cases}
\end{eqnarray}
where
\[\eta_{T} = \left(\frac{L_{\rm tot} \sigma_{T}}{4 \pi m_{p} c^{3} r_{0}}\right)^{\frac{\mu}{1 + 3 \mu}}\]

Here, $L_{\rm tot}$ is the total luminosity available at the base of the jet, and for an approximation $L_{\rm tot} \sim 2 L_{\gamma}$, where $L_{\gamma}$ is the gamma-ray luminosity of the prompt phase \citep{Veres_2012}. $\eta = L/\dot{M}c^2$ is the asymptotic Lorentz factor at the terminal radius \citep{Meszaros_2000}. It has been used as a free parameter, which is equivalent to the bulk Lorentz factor $\Gamma$ for IS and BPH scenarios; however, $\Gamma < \eta$ for the MPH case. The acceleration index $1/3 \leq \mu \leq 1$ ranges between the extreme $\mu = 1/3$ magnetically dominated radial outflow and the usual $\mu = 1$ baryon dominated outflow regimes \citep{M_sz_ros_2011}. Typically, for a magnetically dominated $\mu = 1/3$ case, the photosphere occurs in the acceleration phase $R_{\rm PH} < R_{\rm sat} =  r_0 \eta^{1/\mu}$ with $\eta > \eta_{T}$, the jet accelerates slowly due to magnetic stresses, and $\Gamma(R) \propto R_{PH}^{1/3}$ \citep{B_gu__2015}. On the other hand, for typical baryonic cases, the photosphere occurs in the coasting phase $R_{\rm PH} > R_{\rm sat}$ with $\eta < \eta_{T}$, the jet accelerates linearly in a standard fireball, and $\Gamma(R) \propto R_{\rm PH}$ until the saturation radius \citep{Beloborodov_2010, Gao_2012}.

$\Gamma$ in GRB jets is an important parameter that controls baryon contamination of the jet, neutrino fluence and is useful in the estimation of the emission radius \citep{daigne1998MNRAS_5D, Zhang2013PhRvL01Z}. The value of the emission radius depends on the Lorentz factor, and here we consider the value of $\Gamma$ using empirical correlations between \( \Gamma_0 \) and \( E_{\gamma,\rm iso} \) \citep{Lu_2012ApJ_49L}. The initial Lorentz factor, $\Gamma_0$, is defined during the coasting phase of the outflow, i.e., $\Gamma = \Gamma_0$. 

Using the onset time of the optical afterglow, $\Gamma_0$ has been empirically correlated with the $E_{\gamma, \rm iso}$ of the burst. For a sample of 17 GRBs it was found that $\Gamma_0 \simeq 182 \left(E_{\gamma, \rm iso}/10^{52}~\rm erg\right)^{0.25 \pm 0.03}$ \citep{liang2010ApJ_209L}. This correlation was later refined with larger samples of 51 GRBs, yielding $\Gamma_0 \simeq 91 \left(E_{\gamma, \rm iso}/10^{52}~\rm erg\right)^{0.29}$ \citep{Lu_2012ApJ_49L}. 
In this work, we adopt the latter relation, as it is based on the larger dataset and is therefore statistically more robust.
However, we note that this empirical $\Gamma_0$--$E_{\gamma,\mathrm{iso}}$ correlation exhibits substantial intrinsic scatter when applied to individual bursts. To quantify this uncertainty, we digitized the published sample and evaluated the residual dispersion around the published best-fit relation, obtaining an intrinsic scatter of $\sigma_{\log \Gamma} \simeq 0.27$. This corresponds to a factor of $\sim 2$ uncertainty in the inferred Lorentz factor for individual GRBs. Hence, our derived values those depends on $\Gamma$ are affected by this uncertainty. Recently, the very high energy gamma-ray emission in GRB~221009A has been associated with the presence of cosmic ray acceleration to extreme energies in GRB jets \citep{2022arXiv221012855A, Das2023AnA.._12D}.  At the prompt emission site, due to the interactions of these cosmic rays and intense target photon density, the internal shock and dissipative photosphere model predict higher neutrino flux levels using the photo-hadronic ($p-\gamma$) interaction model \citep{daigne1998MNRAS_5D, rees2005ApJ..847R, rees1994ApJ430L_R, Kobayashi_1997ApJ.IS, albert2017MNRAS_906A, Yangou2024_teccts}. In the $p-\gamma$ process, the delta resonance channel has been opted for our analytical approximations. We assume the proton spectrum follows a power-law distribution $dN_{p}/dE_{p} \propto E_{p}^{-p}$, having a spectral index $p=2$. The target photon distribution is considered as the Band function $n_\gamma(E_\gamma) = dN_\gamma(E_\gamma)/dE_\gamma$ defined as \citep{Band1993ApJ_81B}

\begin{equation}
n_\gamma(E_\gamma) 
= n_{\gamma,b}
\begin{cases}
{\epsilon_{\gamma,b}}^{\alpha}{E_\gamma}^{-\alpha}, & E_\gamma < \epsilon_{\gamma,b} \\[8pt]
{\epsilon_{\gamma,b}}^{\beta}{E_\gamma}^{-\beta},  & E_\gamma \ge \epsilon_{\gamma,b}
\end{cases}
\label{eq:BandBroken}
\end{equation}

where $ n_{\gamma,b}$ is the normalization factor with units $\rm ph/cm^{2}/ keV/s $ at the break energy $\epsilon_{\gamma,b}$, and $\alpha$ and $\beta$ are lower and upper spectral indices.

We have used the model based on the formalism given in \citep{Gao_2013, Zhang2013PhRvL01Z}, where they discuss the neutrino production efficiency for neutrino energy $E_{\nu}$ using the pionization efficiency $f_{\pi}$, which is given by
\begin{align} \label{f_pi} 
f_{\pi} &= 6.13\, \left(\frac{L_{\gamma}}{10^{53} {\rm erg}}\right)\, \left(\frac{R}{10^{13} {\rm cm}}\right)^{-1}\, \left(\frac{\epsilon_{\gamma,b}}{\mathrm{MeV}}\right)\,\left(\frac{\Gamma}{300}\right)^{-2}\, \nonumber\\
&\qquad \times (1 + z)^{-1}\, \left(\frac{E_{\nu}}{E_{\nu,b}}\right)^{\alpha - 1},
\end{align}

The first neutrino break energy $E_{\nu, b}$ occurs due to the threshold of p-$\gamma$ interactions (protons interacting with peak energy of the target photons) at $E_{\nu b, 1} = 6.33 \times 10^{5} (\epsilon_{\gamma, b}/\rm MeV) \ (\Gamma/300)^2 \ (1 + z)^{-2}$ GeV. After the pions decay, the resulting muons also decay into neutrinos. However, these particles can lose energy due to their synchrotron radiation and inverse-Compton (IC) cooling before they decay, which limits the maximum energy of the resulting neutrinos. The second neutrino break energy $E_{\nu b,2}$ is thus set by comparing the pion decay time scale with their synchrotron and IC cooling losses, found to be at $E_{\nu b,2} = 2.12 \times 10^{7} L_{\gamma,53}^{-1/2} \ R_{13} \ \Gamma_{300}^{2} \ (1 + z)^{-1} (\epsilon_{B,-1} + \epsilon_{e,-1})^{1/2} \ \rm GeV$. 
 The charged pion decays to neutrinos via; $\pi^{\pm} \rightarrow \mu^{\pm} + \nu_{\mu} (\bar{\nu}_\mu) \rightarrow \ e^{\pm} + \nu_{e} (\bar{\nu}_e) + \nu_{\mu} + \bar{\nu}_\mu$. The energy of the charged pion is shared among four leptons ($e^{\pm}$, $\nu_{\mu}$, $\bar\nu_{\mu}$ and $\nu_e$ or $\bar\nu_e$) almost equally, that gives $E_{\nu} \simeq E_{\pi}/4$. It is assumed that due to neutrino oscillation, neutrinos of all flavours arrive at Earth in equal numbers. The neutrino flux $\phi$, the rate at which neutrinos pass through a unit area of the IceCube detector \citep{karle2010icecube}, is computed based on the pionization efficiency $f_{\pi}$ as \citep{Gao_2013}:

\begin{multline} \label{n_flux}
\phi(E_{\nu}) = 4.8 \times 10^{-12}~f_p 
\left( \frac{L_\gamma}{10^{53}~\mathrm{erg/s}} \right)^2 
\left( \frac{\epsilon_{\gamma,b}}{\mathrm{MeV}} \right)(1+z)^2 
\left( \frac{\Gamma}{300} \right)^{-6} \\
\times \left( \frac{D_{L}}{10^{27}~\mathrm{cm}} \right)^{-2} 
\left( \frac{R}{10^{13}~\mathrm{cm}} \right)^{-1} 
\left( \frac{E_{\nu}}{E_{\nu b}} \right)^{s-3}, \quad f_\pi < 1 \\[1ex]
\phi(E_{\nu}) = 0.8 \times 10^{-12}~f_p 
\left( \frac{L_\gamma}{10^{53}~\mathrm{erg/s}} \right)(1+z)^3 
\left( \frac{\Gamma}{300} \right)^{-4} \\
\times \left( \dfrac{D_{L}}{10^{27}~\mathrm{cm}} \right)^{-2} 
\left( \dfrac{E_{\nu}}{E_{\nu b}} \right)^{-2},
\quad f_\pi \geq 1 \\
\mathrm{GeV^{-1}\,cm^{-2}}
\end{multline}

where $f_p$ is the baryon loading factor, which describes the fraction of the total GRB energy carried by non-thermal protons, $s$ is the magnitude of high-energy Band index for $E_{\nu} < E_{\nu b,1}$ and magnitude of low energy Band index for $E_{\nu} > E_{\nu b,1}$, and $D_{L}$ is the luminosity distance. Above the second break energy $E_{\nu b,2}$ pion cooling is more important than their decay time scale, and this leads to suppression of the neutrino flux \citep{razza2004PhRv_101R, Gao_2013}. For simplification, beyond $E_{\nu b,2}$ we approximate it by a power law having spectral index $\beta +2$ \citep{Zhang2013PhRvL01Z}.

In the BPH scenario \citep{Pe_er_2007}, the jet carries a large amount of mass, which reduces its ability to accelerate and results in lower Lorentz factors. In contrast, in the MPH scenario \citep{Giannios_2008}, with a low baryon load, the jet is lighter, allowing for efficient acceleration and very high Lorentz factors. In GRB physics, the term baryon loading can carry two related but distinct meanings. In the fireball dynamics, it refers to the ratio \(\eta = L/\dot M c^{2}\), which determines how efficiently the fireball energy is converted into bulk motion and therefore sets the maximum achievable bulk Lorentz factor $\Gamma$ \citep{Gao_2013, Lei_2013}. A heavily baryon-loaded jet (large $\dot M$) results in lower $\Gamma$, while a lightly loaded jet allows for ultra-relativistic expansion. In the context of high-energy neutrino production, baryon loading is instead expressed as the proton energy fraction, \(f_{p} = E_{p}/E_{\gamma, \rm iso}\) \citep{He2012ApJ_.29H, Veres_2024}. Here $E_{p} = \varepsilon_{p} E_{\rm shock}$, and $E_{\gamma, \rm iso} = \varepsilon_{e} f_{e} E_{\rm shock}$; $\varepsilon_{p}$ and $\varepsilon_{e}$ are fractional energies available in protons and electrons at the shock region. $E_{\rm shock}$ is the total energy available in the internal shock or dissipation zone, and $f_{e}$ is the efficiency of electrons radiating their energy as gamma-rays through synchrotron or inverse Compton radiation \citep{He2012ApJ_.29H}. Thus, we obtain the baryon loading $f_{p} = \epsilon_{p}/\epsilon_{e} f_{e}$, which directly governs the normalization of the predicted neutrino flux.

\section{Implications of the IceCube Upper Limits for the GRB Jet}
\label{sec:case_study}

The observations of neutrinos from GRBs are instrumental for investigating the compositions of their jets, the nature of the ejecta, their $\Gamma$, $R$, and microphysical parameters $\epsilon_e$ and $\epsilon_B$, etc. \citep{Zhang2013PhRvL01Z}.  
The time-integrated muon-neutrino fluence in the TeV–PeV energy is used to estimate the normalization of the observed spectrum $\phi_{\nu}(E_{\nu})^{\rm obs} = \phi_0 E_{\nu}^{-2}$. From the time-integrated neutrino fluence upper limit $F_{\nu}^{UL}$, we calculate \(\phi_{0} = F_{\nu}^{UL}/{\rm ln}(E_{\nu, \rm max}/E_{\nu, \rm min})\). The expected number of neutrino events over the neutrino energy range, i.e.,  \(N_{\nu} = \int_{E_{\nu,\rm min}}^{E_{\nu, \rm max}} \ \phi(E_{\nu}) \ A_{\rm eff}(E_{\nu}, \delta) \ dE_{\nu}\), where $A_{\rm eff}(E_{\nu}, \delta)$ IceCube detector's effective area \citep{karle2010icecube}. Further, using our model described in Equations (2)–(4), we constrain $\epsilon_e$, $\epsilon_B$, and explore the allowed parameter space for $\epsilon_p$, $\eta$ and derive lower limit on $R$ and value of $f_p$. We consider $f_e = 0.5$, since typical values of $f_e$ in internal shock and photospheric models are $\le 50 \% $ \citep{Moch1995ApSS_441M,koba1997ApJ_92K,Beni2016MNRAS51B}. Below, we will describe the constraints on the GRB emission models using the IceCube upper limit and explore the model parameter space.

\subsection{GRB~221009A}

GRB~221009A is one of the most energetic GRBs ever observed with an isotropic equivalent gamma-ray energy of \( E_{\gamma,\rm iso} \sim 1.01 \times 10^{55} \,\mathrm{erg} \) \citep{Lesage_2023}. For this GRB, IceCube has provided upper limits on neutrino flux in the energy range MeV to PeV \citep{Abbasi2023ApJ._26A}.
Here, we have used their time-integrated fluence in the TeV-PeV range, based on $T_{90} \simeq 327$ s, and the 90 $\%$ C.L. upper limits are $3.9 \times 10^{-2} \ {\mathrm {GeV~cm^{-2}}}$. Large High Altitude Air Shower Observatory (LHAASO) observation of photons up to 13 TeV from this source indicates the presence of ultra-relativistic protons in the GRB jet \citep{doi:10.1126/sciadv.adj2778, Zhang_B2023ApJ_4Z}. 
Following the relation from \citep{Lu_2012ApJ_49L},  we obtained a Lorentz factor of \( \Gamma_0 \approx 676.5 \). More detailed estimations of the Lorentz factor for GRB~221009A have been discussed in \citep{Lesage_2023}. To optimize the neutrino flux, we have used a value based on \citep{Lu_2012ApJ_49L}, and $\Gamma_0 \simeq 676$ in our calculations. Slightly lower value $\Gamma_0 = 600$ was also reported by \cite{Khan2024ApJ_31K}, for this source.

Based on the location of GRB~221009A provided by the Swift Observatory at the right ascension $\alpha = 288.2645\degree$ and the declination $\delta = +19.7735\degree$ \citep{kruiswijk2024}, we use the IceCube effective area for the range $-5\degree < \delta < 30\degree$ \citep{Aartsen_2017}. Using the IceCube upper limits, we estimate $N_{\nu}$ from this source, and it should be lower than $0.36$ events.

\begin{figure*}[!h]
  \centering
  \begin{subfigure}{0.48\textwidth}
    \centering
    \includegraphics[width=\linewidth]{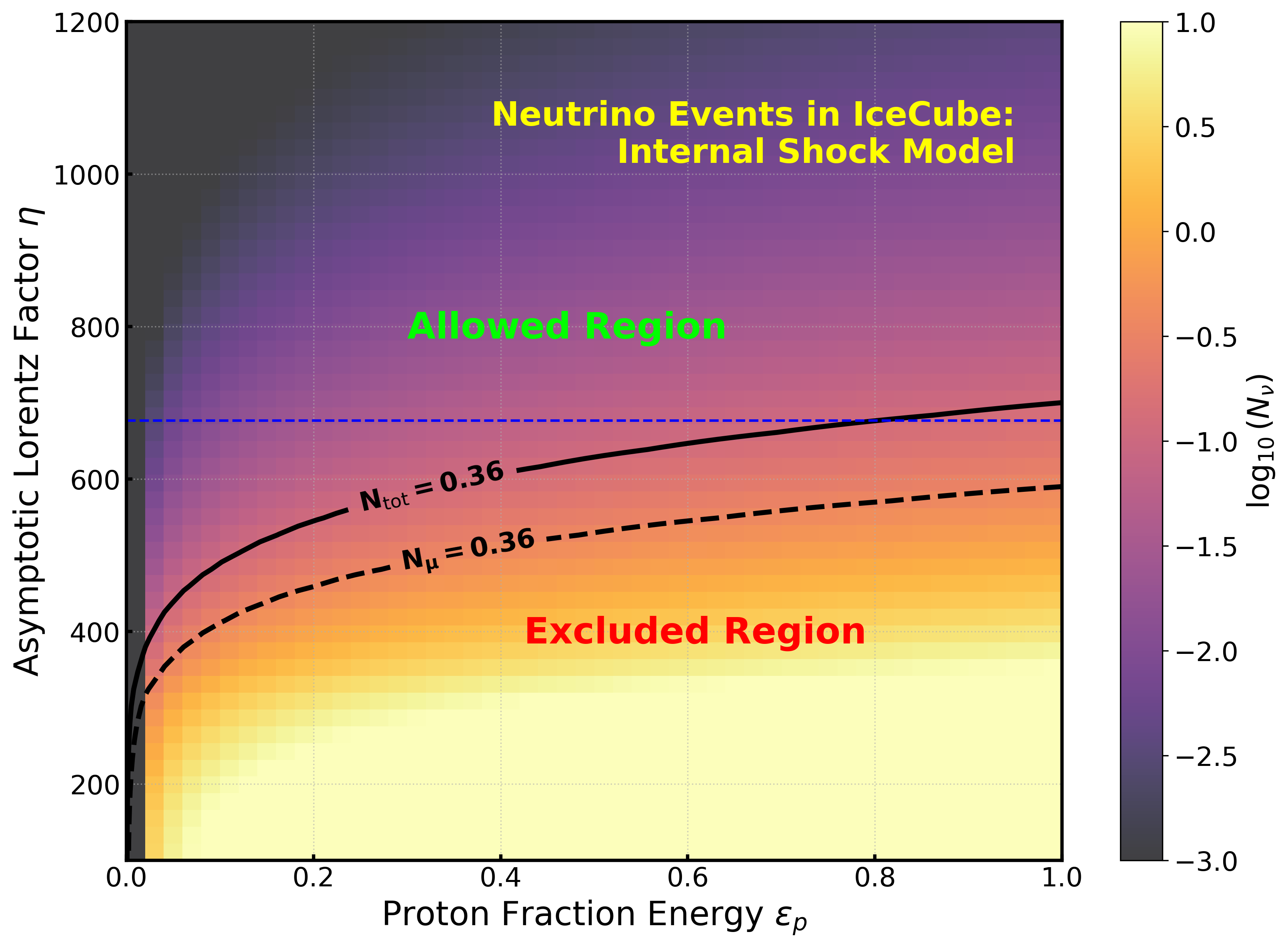}
    \caption{\centering IS Model}
    \label{fig:1_ISmodel_221009A}
  \end{subfigure}\hfill
  \begin{subfigure}{0.48\textwidth}
    \centering
    \includegraphics[width=\linewidth]{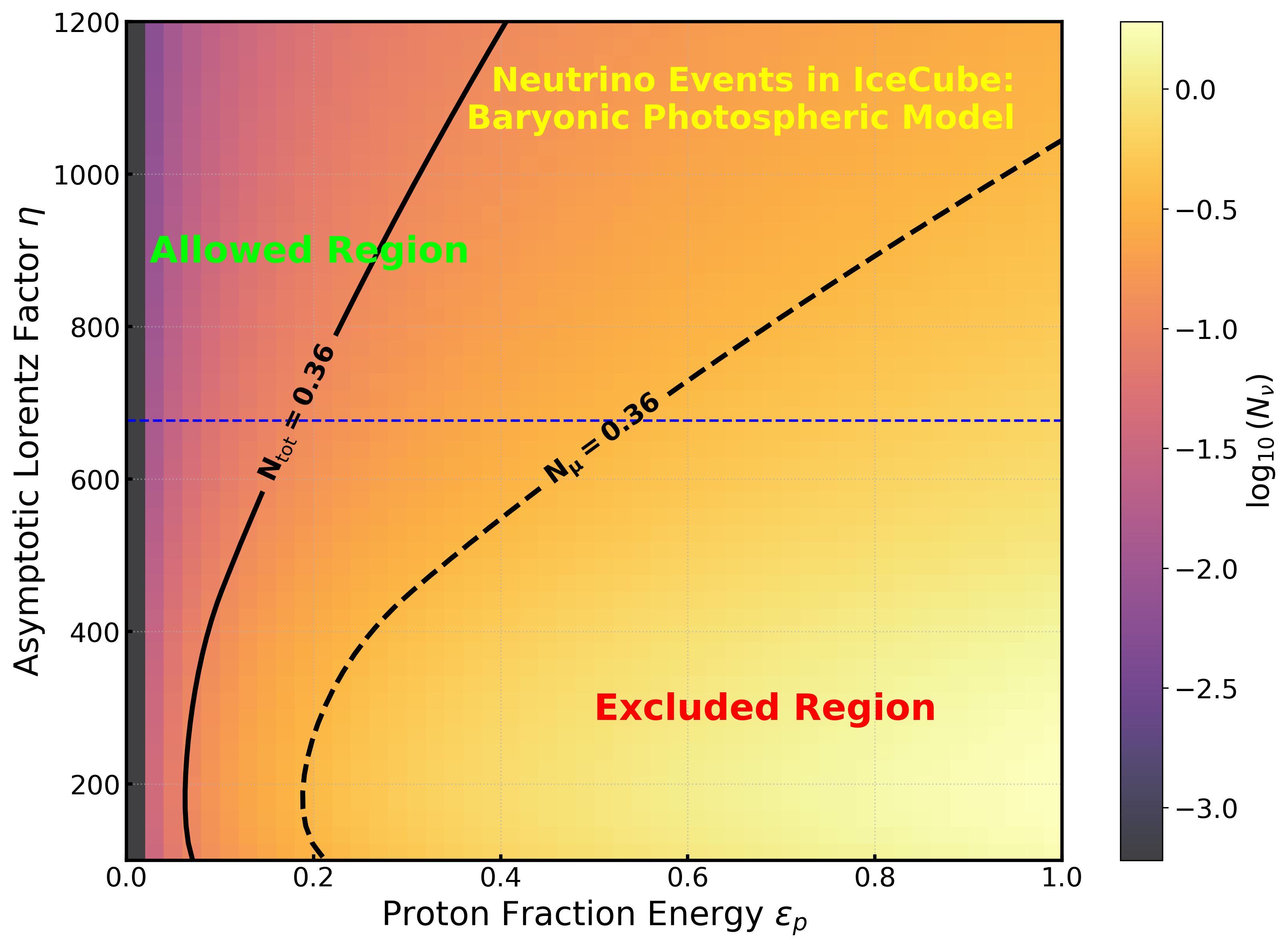}
    \caption{\centering BPH Model}
    \label{fig:BPH_GRB221009A}
  \end{subfigure}\\[1ex]

  \begin{subfigure}{0.48\textwidth}
    \centering
    \includegraphics[width=\linewidth]{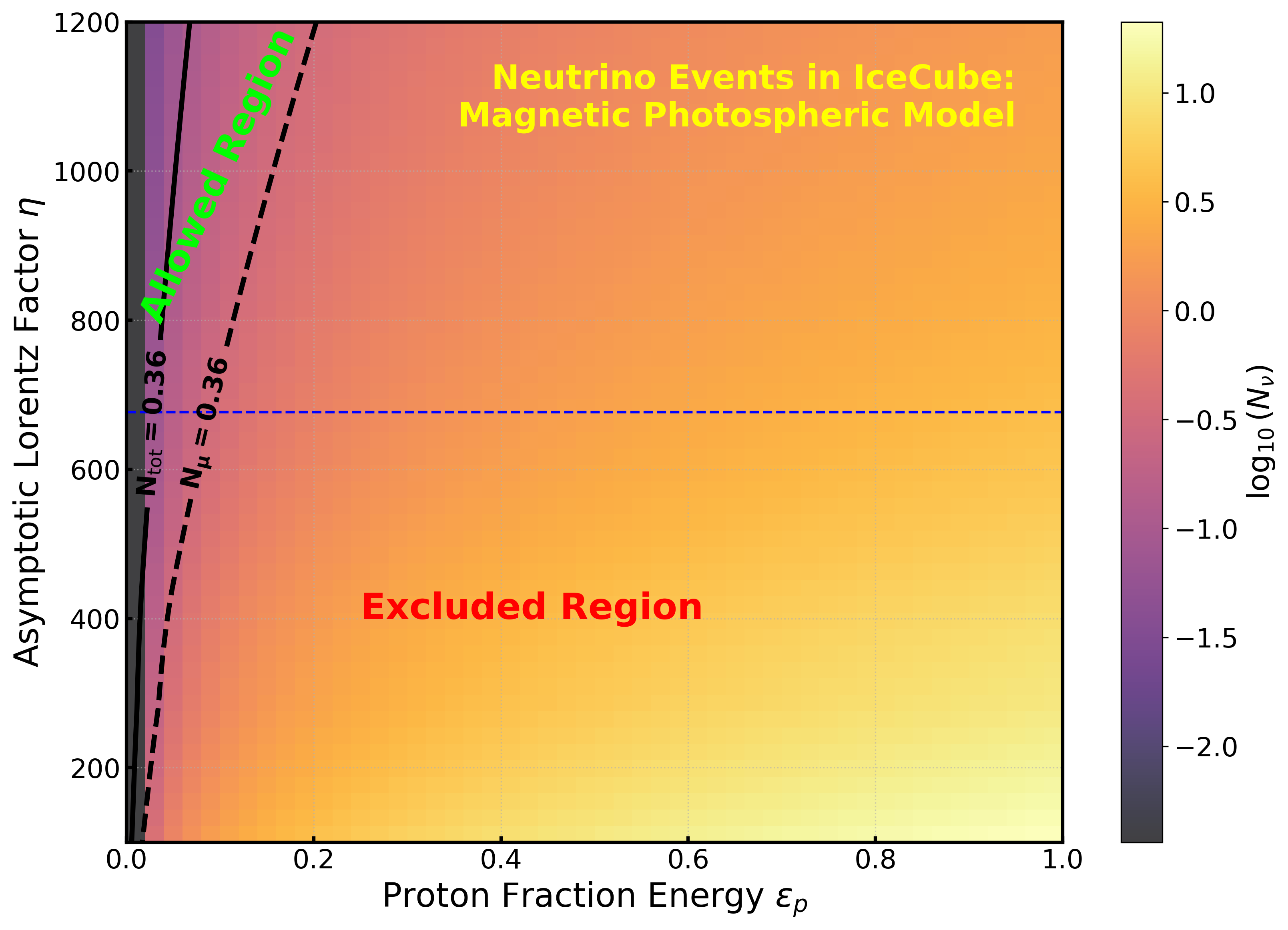}
    \caption{\centering MPH Model}
    \label{fig:MPH_GRB221009A}
  \end{subfigure}\hfill
  \begin{subfigure}{0.48\textwidth}
    \centering
    \includegraphics[width=\linewidth]{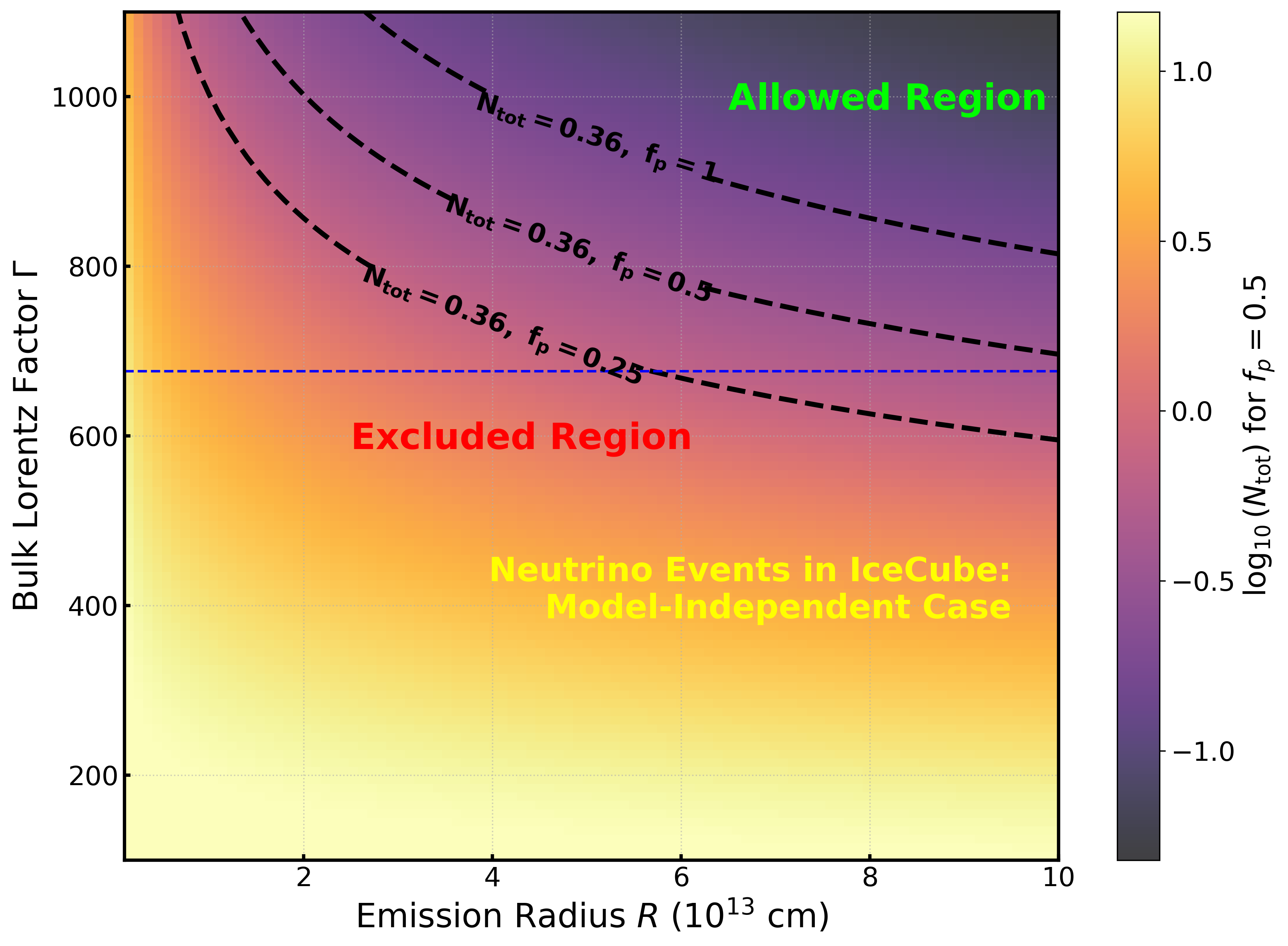}
    \caption{\centering Model Independent}
    \label{fig:Model-Ind_221009A}
  \end{subfigure}

  \captionsetup{width=\linewidth}
  \caption{Density plots of expected neutrino events in IceCube for GRB~221009A. 
  (a) IS Model: The total number $N_{\rm tot}$ (muons + cascades) and $N_{\mu}$ (muons only) are calculated in the two-dimensional parameter space $\eta - \epsilon_{p}$. The colourbar corresponds to $N_{\rm tot}$. Contours where $N_{\rm tot} = 0.36$ and $N_{\mu} = 0.36$ are shown by solid and dashed lines, respectively. The other energy fractions are taken as constants, $\epsilon_{e} = 0.2$ and $\epsilon_{B} = 0.1$. The IceCube null result favours the darker (purple) regions, i.e., high $\eta$ (or $\Gamma$) with weak dependence on $\epsilon_p$. The blue horizontal line marks $\eta = \Gamma = 676$, intersecting the $N_{\rm tot} = 0.36$ contour at $\epsilon_{p} = 0.805$, constraining $f_{p}$. 
  (b) BPH Model: Conventions are the same as (a). A high $\eta$ (or $\Gamma$) is favoured. The line $\eta = \Gamma = 676$ intersects $N_{\rm tot} = 0.36$ at $\epsilon_{p} = 0.18$, constraining $f_{p}$.
  (c) MPH Model: The result is insensitive to $\eta$ (or $\Gamma$) as the contours run almost parallel to the $\eta$-axis. The IceCube null result favours $\epsilon_{p} \lesssim 0.1 - 0.2$, roughly independent of $\eta$. The line $\eta = \Gamma = 676$ intersects $N_{\rm tot} = 0.36$ at $\epsilon_{p} = 0.03$, constraining $f_{p}$.
  (d) Model Independent: Density plot in the parameter space of dissipation radius $R_{13} = (R/10^{13} \rm cm)$ and Lorentz factor $\Gamma$. This uses the semi-analytical model in Section~\ref{IS-PH_Model}, without assuming a specific scenario. Darker colours denote fewer events, while lighter colours indicate more. Dashed lines show contours for $N_{\rm tot} = 0.36$ with $f_{p} = 0.25, 0.5, 1$. Based on the IceCube null result \citep{abbasi2012Natur_I}, the parameter space below each contour is likely ruled out for the corresponding $f_{p}$.}
  \label{fig:all_models_221009A}
\end{figure*}

Using this constraint on $N_{\nu}$, we probe the parameter space $\epsilon_{p}$ and $f_p$. Together, these parameters determine the efficiency of neutrino emission and can be constrained by the observed non-detection. The density plots in Figures \ref{fig:1_ISmodel_221009A}, \ref{fig:BPH_GRB221009A}, and \ref{fig:MPH_GRB221009A} show the allowed and excluded regions of the parameter space for three models: IS, BPH, and MPH, respectively. The dashed line shows the maximum neutrino event for the muon-only flavour, and the solid line represents the sum of all types (track and cascade events) and flavours. These curves represent the upper limits, so the parameter space allowed is towards the lower values with respect to these curves.

In the IS model, as shown in Figure \ref{fig:1_ISmodel_221009A}, $\Gamma = 676$ intersects the exact neutrino events contour line $N_{\rm tot} = 0.36$ at $\epsilon_{p} = 0.805$, and constrains the baryon loading $f_{p} = 8.05$ for $\epsilon_e = 0.2$ and $\epsilon_B = 0.1$. Here, we have used a minimum variability timescale $\Delta t = 0.1 \ s$. A higher value of $\Delta t$ would increase $R_{\rm dis}$ and lower $f_{\pi}$ and $\phi$ according to Equations \ref{f_pi} and \ref{n_flux}, which would lead to a reduction in the production of neutrino events. In the BPH Model, as shown in Figure \ref{fig:BPH_GRB221009A}, $\eta$ increases rapidly, and $\Gamma = 676$ intersects the exact neutrino event contour line $N_{\rm tot}$ at $\epsilon_{p} = 0.18$, and constrains $f_{p} = 1.8$. In the MPH Model (Figure \ref{fig:MPH_GRB221009A}), almost all choices of $\eta$ can be accommodated in the allowed region, which is evident by the contours being almost parallel to the $\eta$-axis as seen in Figure \ref{fig:MPH_GRB221009A}. The value of $\Gamma = 676$ intersects the contour line of the total neutrino events $N_{\rm tot}$ at $\epsilon_{p} = 0.03$, from which we constrain $f_{p} = 0.3$. Figure \ref{fig:Model-Ind_221009A} presents the model-independent study of GRB~221009A. The allowed region in the $R$–$\Gamma$ parameter space favours moderately high $R$ and $\Gamma$ values for low $f_{p}$, while at higher $f_{p}$ only larger $R$ and $\Gamma$ are permitted (as indicated by the darker colour regions).

\subsection{GRB~240825A}

GRB~240825A is a bright long-duration gamma-ray burst with $T_{90} \approx 4\,\mathrm{s}$ and a measured redshift of $z = 0.659$ \citep{gupta_2025arXivG}. The prompt emission was detected across a wide energy range by \emph{Fermi}-GBM, \emph{Fermi}-LAT, and \emph{Swift}-BAT, with an isotropic-equivalent gamma-ray energy of $E_{\gamma,\mathrm{iso}} \approx 2.9\times10^{53}\,\mathrm{erg}$ \citep{2024GCN.37301....1S, 2024GCN.37274....1G}. Spectral analysis of the prompt phase reveals a significant high-energy cutoff at $\sim 79\,\mathrm{MeV}$, which constrains the initial Lorentz factor to $\Gamma_0 < 282$ based on opacity arguments \citep{Wang_2025}. For GRB~240825A, the afterglow onset provides a constraint on the initial Lorentz factor, yielding $\Gamma_{0} \gtrsim 139$ for an ISM environment and $\Gamma_{0} \gtrsim 52$ for a stellar wind environment \citep{cheng2024}. Following the correlation proposed by \citep{Lu_2012ApJ_49L}, the initial Lorentz factor for GRB~240825A is estimated to be $\Gamma_{0} \simeq 242$. This value is consistent with the estimates obtained from other methods for the same burst. In our analysis, we adopt the value $\Gamma_{0} = 242$.

\begin{figure*}[!h]
  \centering
  \begin{subfigure}{0.48\linewidth}
    \centering
    \includegraphics[width=\linewidth]{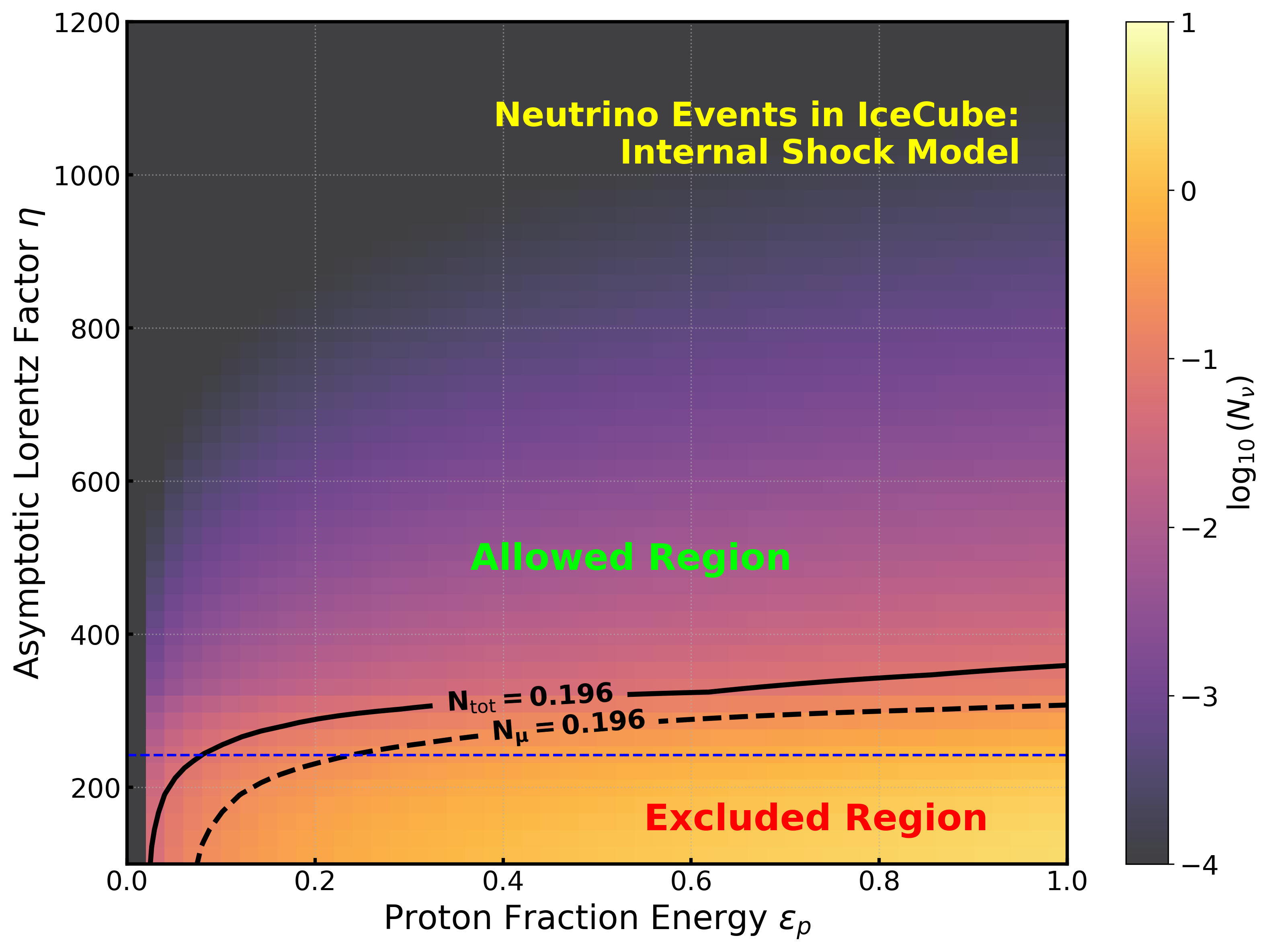}
    \caption{\centering IS Model}
    \label{fig:1_ISmodel_240825A}
  \end{subfigure}\hfill
  \begin{subfigure}{0.48\linewidth}
    \centering
    \includegraphics[width=\linewidth]{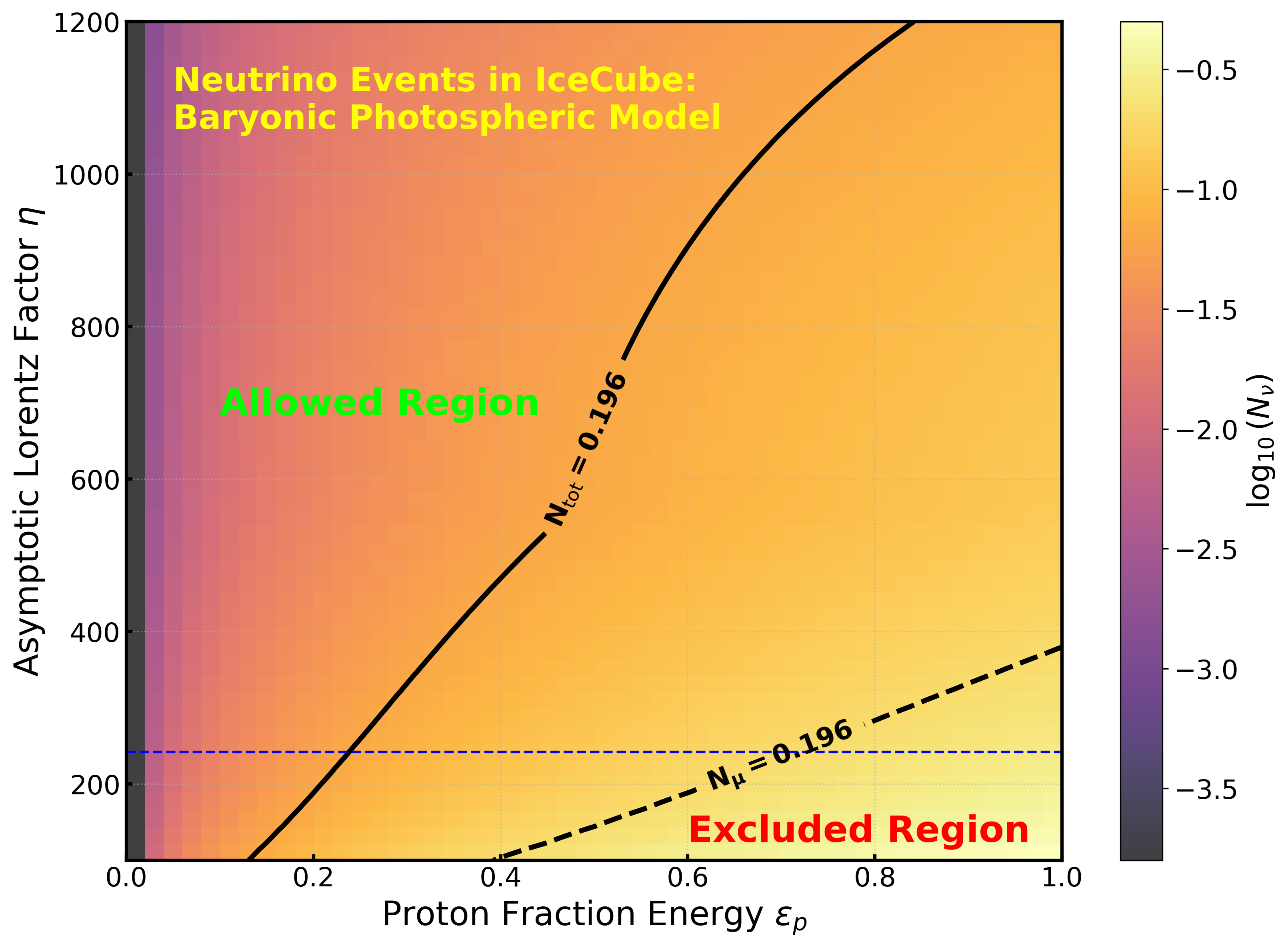}
    \caption{\centering BPH Model}
    \label{fig:BPH_GRB240825A}
  \end{subfigure}\\[1ex]

  \begin{subfigure}{0.48\linewidth}
    \centering
    \includegraphics[width=\linewidth]{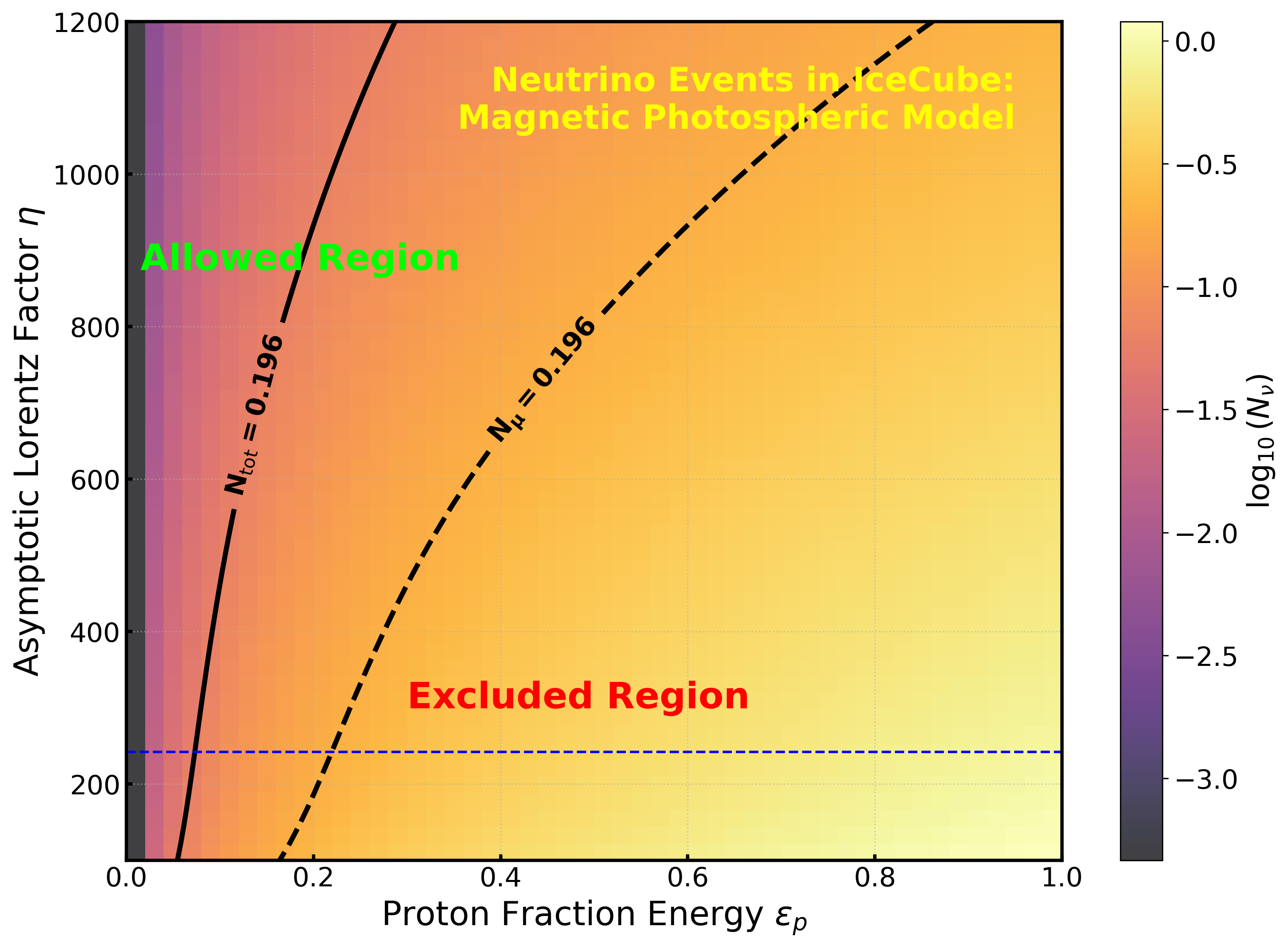}
    \caption{\centering MPH Model}
    \label{fig:MPH_GRB240825A}
  \end{subfigure}\hfill
  \begin{subfigure}{0.48\linewidth}
    \centering
    \includegraphics[width=\linewidth]{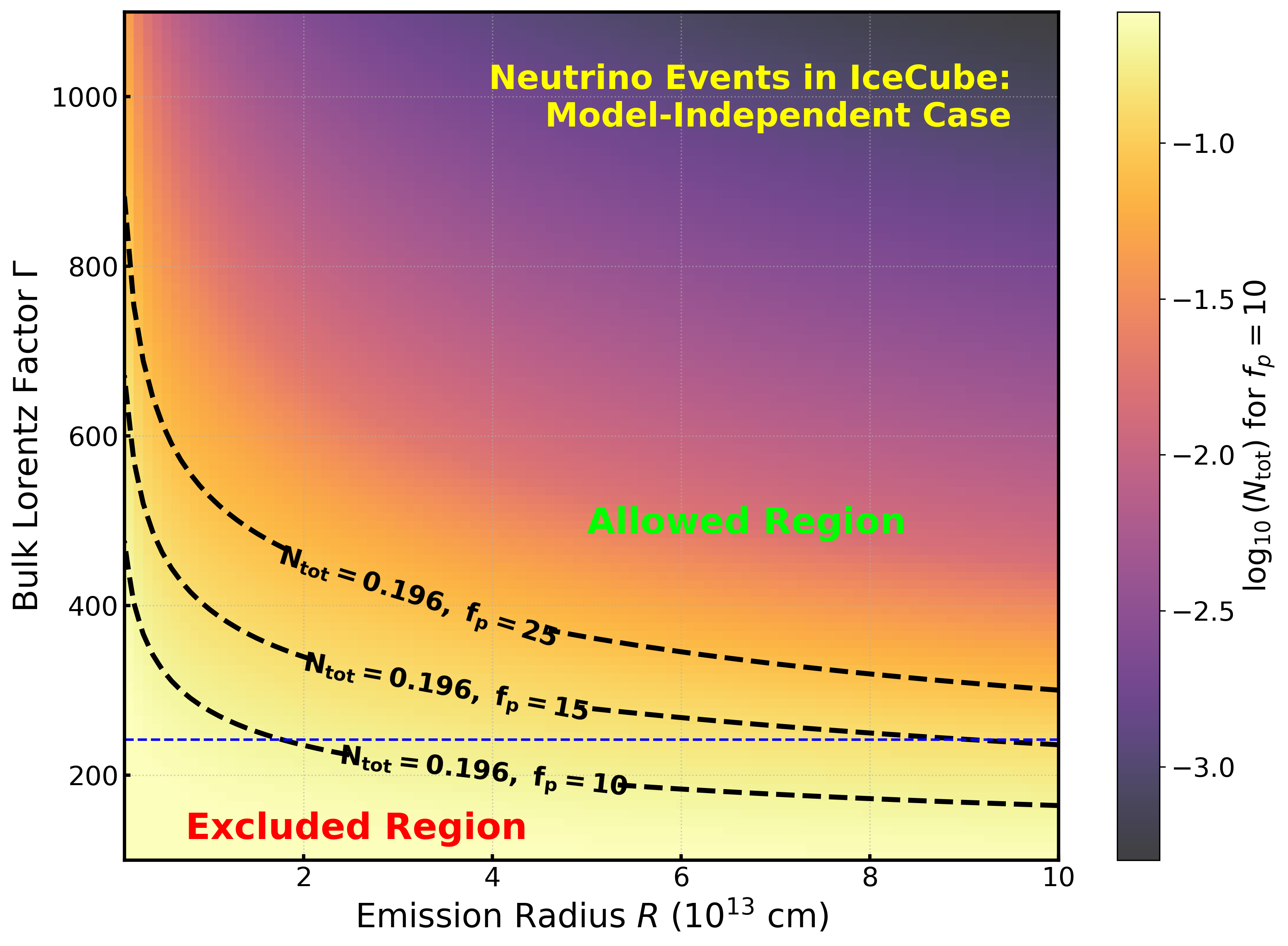}
    \caption{\centering Model Independent}
    \label{fig:Model-Ind_240825A}
  \end{subfigure}

  \captionsetup{width=\linewidth}
  \caption{Density plots of expected neutrino events in IceCube for GRB~240825A. 
  (a) IS Model: Contours where $N_{\rm tot} = 0.196$ and $N_{\mu} = 0.196$ are shown by solid and dashed lines, respectively. The blue horizontal line marks $\eta = \Gamma = 242$, intersecting the $N_{\rm tot} = 0.196$ contour at $\epsilon_{p} = 0.08$, constraining $f_{p}$. The other energy fractions are fixed to $\epsilon_{e} = 10^{-2}$ and $\epsilon_{B} = 10^{-3}$.
  (b) BPH Model: Conventions are the same as (a). The blue line $\eta = \Gamma = 242$ intersects $N_{\rm tot} = 0.196$ at $\epsilon_{p} = 0.238$, setting the $f_{p}$ upper bound.
  (c) MPH Model: Results are insensitive to $\eta$, as contours run nearly parallel to the $\eta$-axis. IceCube non-detection favours $\epsilon_{p} \lesssim 0.1$ -- $0.3$, weakly dependent on $\eta$. The $\eta = 242$ line intersects $N_{\rm tot} = 0.196$ at $\epsilon_{p} = 0.073$.
  (d) Model Independent: Density plot in the dissipation radius $R_{13} = (R_{\rm dis}/10^{13})$ cm vs.\ $\Gamma$. Darker colours denote fewer events. Dashed lines show $N_{\rm tot} = 0.196$ contours for $f_{p} = 10, 15, 25$.}
  \label{fig:all_models_240825A}
\end{figure*}

Following the detection of GRB~240825A (R.A. = 341.6\degree, $\delta = 5.9\degree$), the IceCube Neutrino Observatory conducted a targeted search for track-like muon-neutrino events in temporal coincidence with the burst. No significant excess above background was found. For an assumed $E^{-2}$ spectrum, the 90\% confidence-level upper limit on the time-integrated muon-neutrino fluence in the TeV–PeV energy range is $2.8\times10^{-2}~\mathrm{GeV\,cm^{-2}}$ \citep{2024GCN.37326....1I}, for the prompt window defined as $T_0 - 1\,\mathrm{h}$ to $T_0 + 2\,\mathrm{h}$. When extending the search to a $\pm 1$ day window around the trigger time, the corresponding upper limit becomes $3.0\times10^{-2}~\mathrm{GeV\,cm^{-2}}.$ We calculated the expected number of neutrino events for the  GRB~240825A corresponding to this observational upper limit on the neutrino flux by convolving the flux upper limit with the IceCube effective area for the range $-5\degree < \delta < 30\degree$ \citep{Aartsen_2017}, similar to that of GRB~221009A. The value of $N_{\nu}$ in this case would be below $0.196$ events. We find $\epsilon_{e} = 10^{-2}$, $\epsilon_{B} = 10^{-3}$, which are factors of 10 and 100 lower, respectively, than the values for GRB~221009A.

In the IS model (Figure~\ref{fig:1_ISmodel_240825A}), unlike GRB~221009A, where $\epsilon_{p}$ was found to have a very high value, the case of $\Gamma = 242$ intersects the neutrino event contour line $N_{\rm tot} = 0.196$ at $\epsilon_{p} = 0.08$. This leads to a moderate value of $f_{p} = 16$. In the BPH model (Figure~\ref{fig:BPH_GRB240825A}), the value of $\Gamma = 242$, intersects $N_{\rm tot}$ contour at $\epsilon_{p} = 0.238$, indicating a much higher baryon loading of $f_{p} = 47.6$. Meanwhile, in the MPH model (Figure~\ref{fig:MPH_GRB240825A}), $\epsilon_{p} = 0.073$ corresponds to $f_{p} = 14.6$. The model-independent analysis of GRB~240825A (Figure~\ref{fig:Model-Ind_240825A}) imposes stringent constraints on the $R_{\rm dis}$–$\Gamma$ parameter space, particularly at large baryon loadings, thereby favouring moderately high values of both $R_{\rm dis}$ and $\Gamma$. Since for GRB~240825A we adopt $\epsilon_{e} = 10^{-2}$ and $f_{e} = 0.5$, the corresponding baryon loading becomes $f_{p} = 20$ for the model independent case.

\section{Prompt Emission Site via $\gamma-\gamma$ Opacity}
\label{sec:Optical_depth}
The maximum energy of the photons observed during the prompt emission phase is useful to estimate the emission radius of gamma-ray photons \citep{Gupta_2008MNRASG, Zhang_2009, Chand_2020}. However, the maximum energy of the photons connected with the prompt phase is model-dependent. For example, in GRB~130427A, it was shown that the maximum photon energy 73 GeV was associated with the prompt emission via photo-disintegration of iron nuclei \citep{jagdish2016_9J} but other models associated it with the afterglow phase in GRBs via synchrotron self-Compton scenario \citep{2013ApJ_..20L}. In the LHAASO collaboration paper, for GRB~221009A, they have assumed that TeV photons are produced in the prompt phase and hence higher values of opacity $\tau_{\gamma\gamma} \simeq 190$ are reported \citep{2023Sci...380.1390L}. However, these TeV photons may also originate in the afterglow region in this source \citep{Foffano2024ApJ_44F,2025MNRAS.540.2098K}. Hence, the ambiguity still remains for the highest energy photons that originated close to the end of the prompt phase.

In case of GRB~240825A for the prompt phase, we have the observed spectral cutoff at $\sim 79\,\mathrm{MeV}$ \citep{Wang_2025}. In contrast, for GRB~221009A the cutoff energy is $\sim 100\,\mathrm{MeV}$ \citep{Axelsson2025A_4A}. For these photons, we estimate the emission radius using the pair production optical depth $ \tau_{\gamma\gamma}(E)$ of photons of energy $E$ in the radiation field of target photons using the formalism discussed in \citep{Gupta_2008MNRASG, Zhang_2009, Gould1967PhRv_4G}:

\begin{figure}[h]
    \centering
    \includegraphics[width=\linewidth]{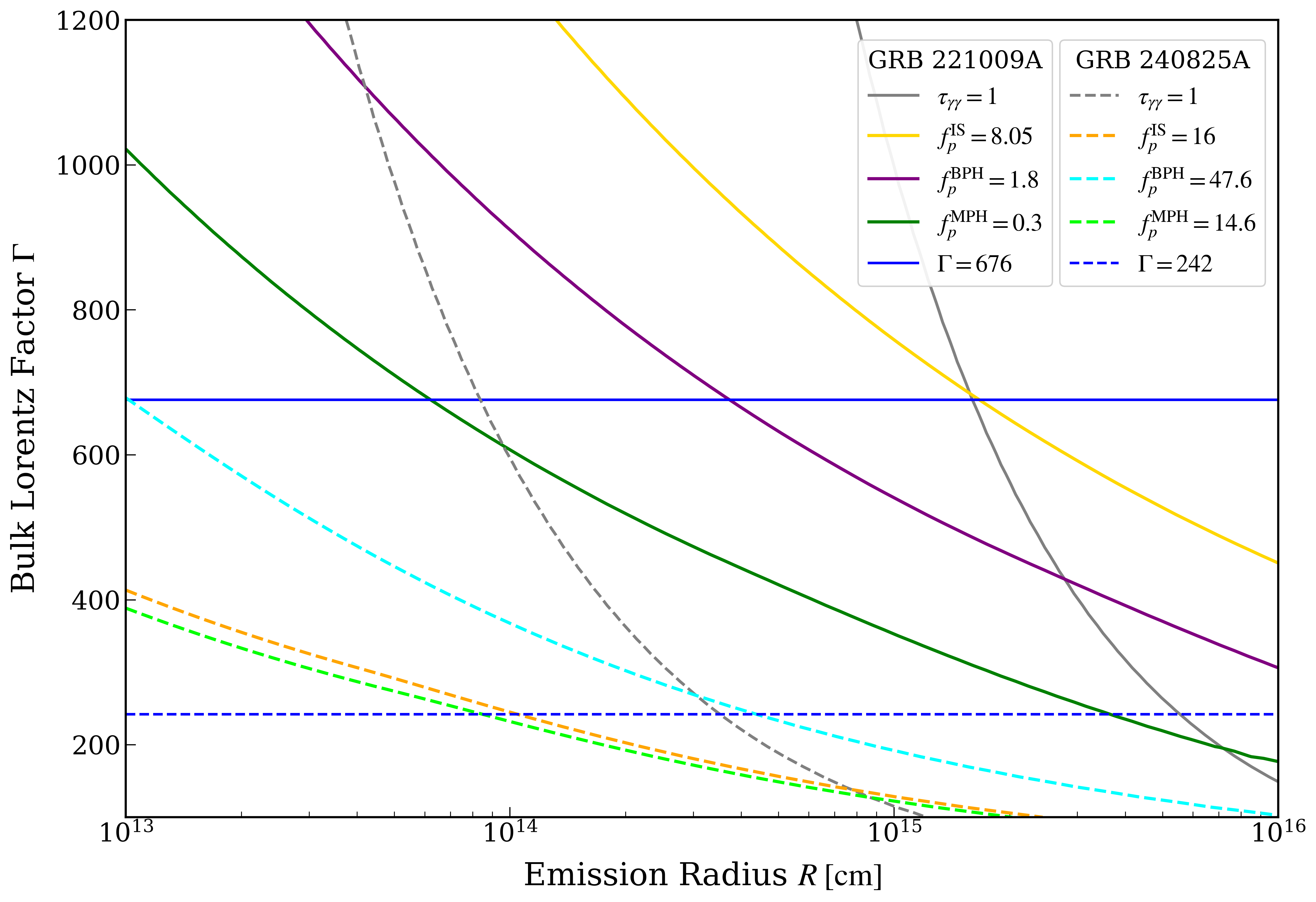}
    \captionsetup{width=\linewidth}
    \caption{$R_{\rm dis}-\Gamma$ curves for various models based on the baryon loading values. These curves are compared with the $\tau_{\gamma\gamma} =1$ curve for GRB~221009A and GRB~240825A. }
    \label{fig:comp_surface}
\end{figure}

\begin{align}
    \tau_{\gamma\gamma}(E) = \frac{C(\beta) \sigma_{T} d_{z}^{2} f_{0}}{-1 -\beta} \left(\frac{E}{m_{e}^{2} c^{4}}\right)^{-1-\beta} \frac{1}{R^{2}} \left(\frac{\Gamma}{1 + z}\right)^{2 + 2 \beta}
\end{align}

where $C(\beta) \simeq (7/6) (-\beta)^{-5/3}/ (1-\beta)$ \citep{Svensson1987MNRAS}, $\sigma_T$ is the Thomson cross section, $m_e$ is the electron mass, $c$ is the speed of light, $d_z$ is the co-moving distance and $\beta$ is the spectral index of the power law spectrum of the Band function after the peak energy \citep{Band1993ApJ_81B}, $f_0$ is parameterized in units of $\rm ph \ cm^{-2} \ (keV)^{-1-\beta}$, and its value is calculated based on the formalism given in \citep{Abdo2009Sc_88A}.

We have shown the electromagnetically thin $R-\Gamma$ curve by assuming the condition $\tau_{\gamma\gamma} = 1$ during the prompt phase in Figure \ref{fig:comp_surface} by a grey solid and dashed line for GRB~221009A and GRB~240825A, respectively.  We also show lower limits on the dissipation radius using neutrino upper limits in Figure \ref{fig:comp_surface}. This indicates that dissipation might occur in optically thin regions, which is consistent with the non-detection of neutrino events. We note that our opacity estimates are based on a simplified one-zone, time-independent treatment, and do not account for angular or temporal dependencies. In particular, for GRB~221009A, there is evidence for a structured jet geometry and extended high-energy emission \citep{gill2023MNRAS_8G, Ren2024ApJ_115R, Mondal2025_4893M}, which may reduce the effective photon density at late times or for off-axis emission. Hence, our opacity estimates may be conservative and could overestimate the true opacity for the highest-energy photons.

\begin{table}[!htbp]
\centering
\resizebox{\columnwidth}{!}{
\begin{tabular}{lcc}
\hline
Model Parameters & GRB~221009A & GRB~240825A \\
\hline
 & Observable Parameters  & \\
\text{z} & 0.151 [1] & 0.659 [2] \\
\text{$E_{\rm peak}$ [MeV]} & 2.66 [3] & 0.502 [4] \\
\text{$\epsilon_{\gamma,b}$ [MeV]} & 0.92 [5] & 0.375 [3] \\
\textbf{$\alpha$} & -0.89 [6] & -0.9 [4] \\
\textbf{$\beta$} & -2.21 [6] & -2.4 [4] \\
\text{$L_{\gamma,\rm iso}$ [erg/s]} & $9.9 \times 10^{53}$ [1] & $3.04 \times 10^{53}$ [7] \\
\text{$E_{\gamma,\rm iso}$ [erg]} & $1.01 \times 10^{55}$ [1] & $2.93 \times 10^{53}$ [2] \\
\text{$D_{L}$ [cm]} & $2.233 \times 10^{27}$ [1] & $12.355 \times 10^{27}$ [2] \\
\text{$T_{90}$ [s]} & 327 [8] & 4 [7] \\
 & Model Parameters  & \\
\textbf{$\Gamma$} & 676   & 242 \\
\text{$\delta t_{min}$ [s]} & $0.1$ & 0.058 \\
\textbf{$\epsilon_{e}$} & 0.2 & $10^{-2}$ \\
\textbf{$\epsilon_{B}$} & 0.1 & $10^{-3}$ \\
 & Derived Parameters  & \\
\multirow{3}{*}{$R$ [cm]} 
  & $\gtrsim 2.386 \times 10^{15}$ (IS) & $\gtrsim 1.228 \times 10^{14}$ (IS) \\
  & $\gtrsim 1.876 \times 10^{12}$ (BPH) & $\gtrsim 1.259 \times 10^{13}$ (BPH) \\
  & $\gtrsim 3.127 \times 10^{11}$ (MPH) & $\gtrsim 3.147 \times 10^{12}$ (MPH) \\
\multirow{3}{*}{$f_{p}$} 
  & 8.05 (IS)   & 16 (IS) \\
  & 1.8 (BPH) & 47.6 (BPH) \\
  & 0.3 (MPH) & 14.6 (MPH) \\
\hline
\end{tabular}
}
\caption{GRB Parameters: The parameters, in the same order as mentioned, are redshift, peak energy of the observed $\gamma$-ray fluence, photon break energy, low- and high-energy index of the Band function, isotropic luminosity, isotropic energy, luminosity distance, $T_{90}$-duration, bulk Lorentz factor as calculated by \citep{Lu_2012ApJ_49L}, cut-off Lorentz factor, minimum variability timescale, electron and magnetic field fraction energy, lower limit on the emission radius, and baryon loading in IS, BPH and MPH models (as calculated analytically), respectively. In our work, we have chosen $\delta t_{min} = 0.1 \ s$ for GRB~221009A, while $\delta t_{min} = 0.058 \ s$ for GRB~240825A \citep{2024GCN.37302....1F}.\\
-----------------------------------------------------------------------------------------------
References: [1] \citep{Lesage_2023}, [2] \citep{Zhang_2025}, [3] \citep{GRB221009A_Frederiks}, [4] \citep{Wang_2025}, [5] \citep{GRB221009A_Frederiks}, [6] \citep{GRB221009A_Frederiks}, [7] \citep{2024GCN.37302....1F}, [8] \citep{Abbasi2023ApJ._26A}.}
\label{tab:GRBP}
\end{table}

\section{Results and Discussion}
\label{sec:results}

We study prompt emission models (IS, BPH, and MPH) that are promising for producing neutrino flux via photo-hadronic interactions. We consider a Lorentz factor of $\Gamma = 676$ for GRB~221009A. We estimate our constraints on $f_p$ by assuming $f_e = 0.5$. First, we constrain the values of the microphysical parameters, $\epsilon_e = 0.2$ and $\epsilon_B = 0.1$, using the IceCube upper limits, as described in Figure \ref{fig:all_models_221009A}, for GRB~221009A. Further, using the density plots for this source, we constrain $f_{p} \ \leq 8.05$ for the IS model. However, for the BPH model this value is $f_{p} \ \leq 1.8$ and for the MPH model $f_{p} \ \leq 0.3$. The corresponding estimated dissipation radii for the prompt phase in different radiation models are $R_{\rm MPH} \gtrsim 3.1 \times 10^{11}$ cm, $R_{\rm BPH} \gtrsim 1.8 \times 10^{12}$ cm, and $R_{\rm IS} \gtrsim 2.4 \times 10^{15}$ cm, respectively. In the IS Model, the value of $\epsilon_{p} = 0.805$, i.e., $f_{p} = 8.05$, coincides with the upper limit as seen with the solid-dashed lines plotted for $0.36$ neutrino events. In the BPH Model, the value of $\epsilon_{p} = 0.18$, i.e., $f_{p} = 1.8$, coincides with $0.36$ neutrino events. In the MPH Model, when the magnetization parameter $\sigma >> 1$, the maximum value of $\epsilon_{p} = 0.03$, i.e., $f_{p} = 0.3$, coincides with the upper limit of $0.36$ neutrino events. Since the MPH model predicts strong magnetization, it follows that $\epsilon_{p}$ will always remain below $0.1$.

We follow a similar approach for GRB~240825A by considering $\Gamma = 242$. In this case, the values of the microphysical parameters are $\epsilon_e = 10^{-2}$ and $\epsilon_B = 10^{-3}$. The baryon loading for the IS model is $f_{p} \ \leq 16$, for the BPH model $f_{p} \ \leq 47.6$, and $f_{p} \ \leq 14.6$ for the MPH model. The dissipation radius values are $R_{\rm MPH} \gtrsim 3.2 \times 10^{12}$ cm, $R_{\rm BPH} \gtrsim 1.3 \times 10^{13}$ cm and $R_{\rm IS} \gtrsim 1.2 \times 10^{14}$ cm, respectively. Further, the IS Model gives the maximum value of $\epsilon_{p} = 0.08$, i.e., $f_{p} = 16$, which coincides with the upper limit as seen with the solid-dashed lines plotted for $0.196$ neutrino events. In the BPH Model, the maximum value of $\epsilon_{p} = 0.238$, i.e., $f_{p} = 47.6$, coincides with $0.196$ neutrino events. In the MPH Model, when the magnetization parameter $\sigma >> 1$, the maximum value of $\epsilon_{p} = 0.073$, i.e., $f_{p} = 14.6$, coincides with $0.196$ neutrino events. These values are consistent with a matter-dominated (low-magnetization) jet in the framework of the internal shock model of GRBs \cite{Bos2009A_77B, Beni2013ApJ_69B}. An important caveat of our analysis concerns the coupling between $\Gamma$, $f_p$ and $R$ in neutrino production models. Neutrino emission is primarily sensitive to combinations of these parameters. By fixing $\Gamma$ using an empirical correlation, we effectively reduce this parameter degeneracy, and the resulting constraints on $f_p$ and R should therefore be interpreted as conditional limits rather than absolute measurements. A full exploration of the coupled $(\Gamma, f_p, R)$ parameter space, which would require a joint multi-dimensional analysis, is beyond the scope of the present work and is left for future studies. In earlier studies, for $f_p \lesssim 10$, the constraint on the dissipation radius was reported to be $R \gtrsim 4 \times 10^{12}$ cm \citep{He2012ApJ_.29H, Li_2012PhRvD7301L}. We also found $f_p < 10$ for GRB~221009A, whereas for GRB~240825A, its value is $47.6$ for the BPH case.

\begin{figure*}[!htbp]
  \centering
  \begin{subfigure}{1.0\linewidth}
    \centering
    \includegraphics[width=\linewidth]{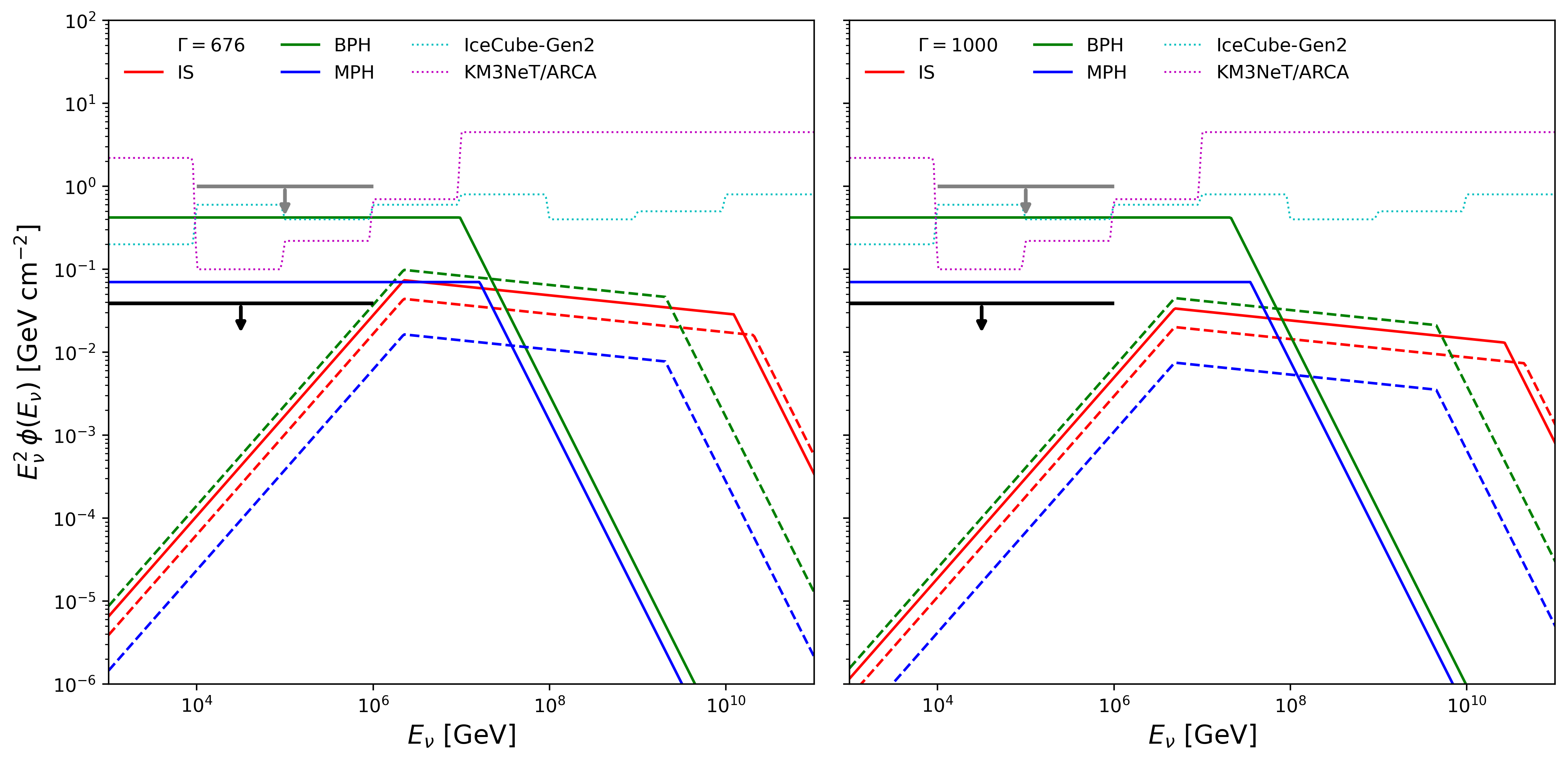}
    \caption{\centering Neutrino Fluence of GRB~221009A based on the photo-hadronic interaction model}
    \label{fig:IS-BPH-MPH_GRB221009A}
  \end{subfigure}

  \begin{subfigure}{1.0\linewidth}
    \centering
    \includegraphics[width=\linewidth]{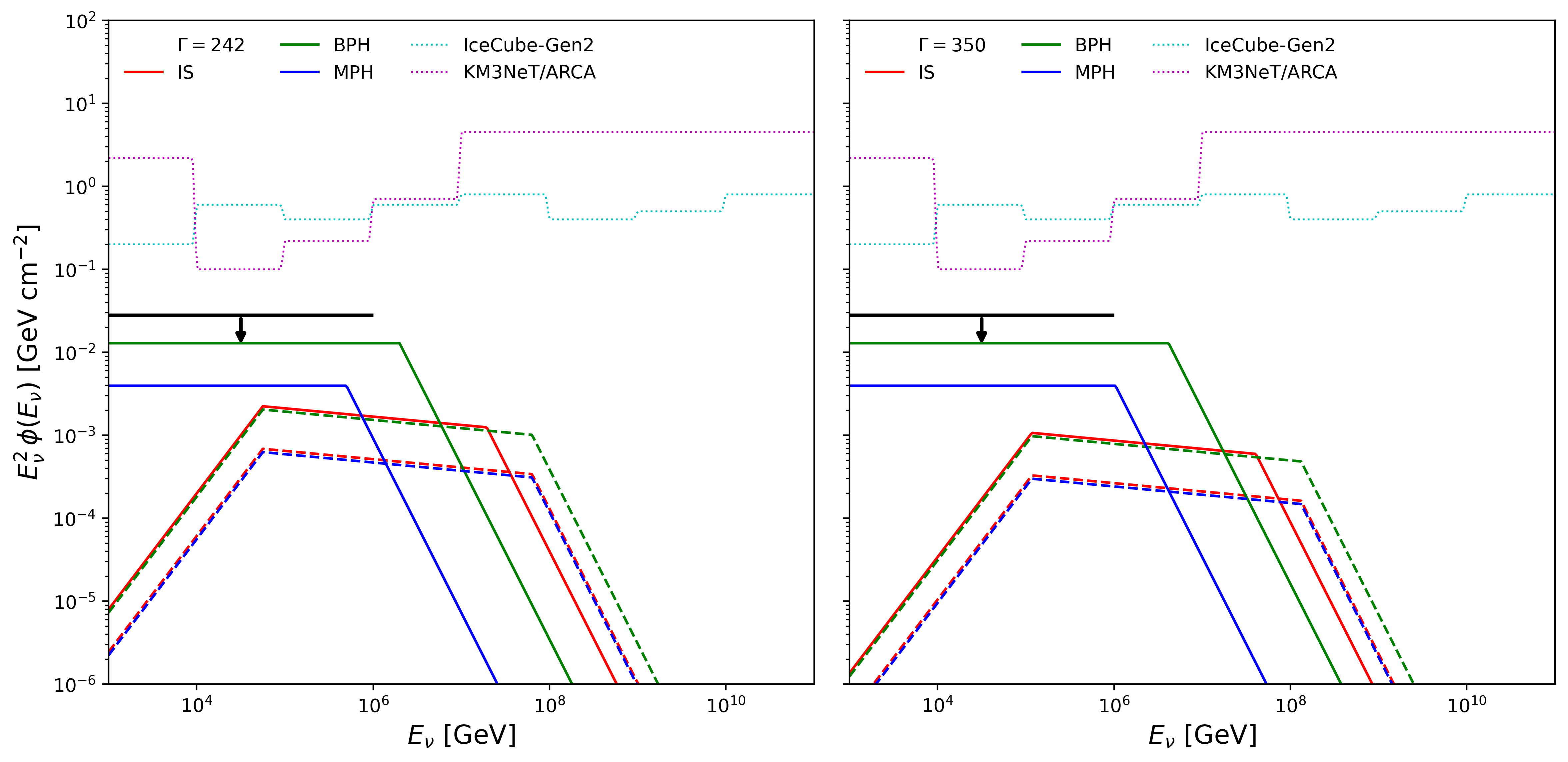}
    \caption{\centering Neutrino Fluence of GRB~240825A based on the photo-hadronic interaction model}
    \label{fig:IS-BPH-MPH_GRB240825A}
  \end{subfigure}

  \captionsetup{width=\linewidth}
  \caption{Predicted neutrino spectrum of GRB~221009A and GRB~240825A for the IS, BPH, and MPH models, along with IceCube-Gen2 ($\pm 500s$ time window) \citep{aartsen2021JPhG_A} and KM3NeT/ARCA (1000s time window) sensitivities \citep{Palacios-Gonzalez:2021slw}. The IceCube and KM3NET/ARCA upper limits are shown with the black and grey horizontal lines, respectively, for each GRB.
  (a) GRB~221009A: For $\Gamma = 676$ and $1000$, solid lines represent the value of dissipation radius as mentioned in Table \ref{tab:GRBP}, while the dashed lines represent $R = 4 \times 10^{15} $ cm for IS  model, and $R = 4 \times 10^{14}$ cm for BPH and MPH models; the baryon loading values are $f_{p} = 8.05$ (IS), $1.8$ (BPH), and $0.3$ (MPH).
  (b) GRB~240825A: For $\Gamma = 242$ and $350$, solid lines represent the value of dissipation radius as mentioned in Table \ref{tab:GRBP}, while the dashed lines represent $R=4 \times 10^{14} $ cm for all models; the baryon loading values are $f_{p} = 16$ (IS), $47.6$ (BPH), and $14.6$ (MPH), respectively.   \label{fig:fluence_comparison}}
\end{figure*}

In Figure \ref{fig:comp_surface}, we have shown the $R-\Gamma$ parameter space based on the $f_{p}$ values estimated for both GRBs using IceCube upper limits. We further notice that dissipation radii for various models are lower than the $\tau_{\gamma\gamma} =1$ curve (grey solid line for GRB~221009A, and grey dashed line for GRB~240825A), where pair production opacity is higher due to higher target photon density for pair production. The neutrino fluence associated with GRB~221009A is shown in Figure \ref{fig:IS-BPH-MPH_GRB221009A} for $\Gamma = 676$ and 1000 in the left and right panels.  Similarly, in Figure \ref{fig:IS-BPH-MPH_GRB240825A}, neutrino fluence for the models is shown for $\Gamma = 242$ and $350$ in the left and right panels, respectively. In both figures, the solid curves are shown for the minimum value of $R$ as listed in Table 1. The dashed curves for GRB~221009A are plotted at a larger value of $R = 4 \times 10^{15}$ cm for the IS model, and $R = 4 \times 10^{14}$ cm for both BPH and MPH models. In the case of GRB~240825A, we have taken $R = 10^{14}$ cm for all dashed curves. We consider the upper limits of IceCube and KM3NeT/ARCA, which are available in an energy band, and the constraints could be relaxed outside the energy range, similar to that in \citep{AI_2023ApJ_115A}.

In the light of current TeV-PeV neutrino upper limits for GRB~240825A, all models are possible. However, for GRB~221009A, BPH and MPH models can be discarded for the minimum value of $R$ for both values of $\Gamma = 676$ and $1000$, and the IS model is a feasible one. For our increased value of $R$, all models are consistent with the IceCube upper limits. Neutrino detection with the IceCube Gen-2 and KM3NeT sensitivities for any future GRB with similar properties as GRB~240825A would still be difficult, as the neutrino signal remains lower than the IceCube Gen-2 and KM3NeT sensitivities. Future generation telescopes, like Giant Radio Array for Neutrino Detection (GRAND) and IceCube-Gen2, would be more appropriate for neutrino detection from individual GRBs and reveal the complex nature of GRB jets \citep{aartsen2021JPhG_A, alvarez2020SCPMA_A, zhang2025ApJ_9Z}.

\section{Acknowledgements}

The authors thank S. Sahu and T. A. Dzhatdoev for reading the manuscript and for their comments and S. Razzaque, V. Chand for discussions. We also thank the referee for constructive suggestions that improved the paper. KS acknowledges financial support from the Innovation in Science Pursuit for Inspired Research (INSPIRE) Fellowship, Department of Science \& Technology, Government of India (DST/INSPIRE Fellowship/2021/IF210721).


\newpage
\onecolumn
\bibliographystyle{mnras}
\bibliography{refer_rgamma}

\end{document}